\documentclass[prl,twocolumn,superscriptaddress]{revtex4-1}

\usepackage{amsmath,amssymb,amsfonts,bm,color,graphicx,tabularx,physics}

\usepackage[unicode=true,colorlinks=true]{hyperref}

\usepackage[etex=true,export]{adjustbox}

\hypersetup{linkcolor=blue,citecolor=blue,urlcolor=blue}

\newcommand{\beq}{\begin{equation}}
\newcommand{\eeq}{\end{equation}}
\newcommand{\beql}{\begin{equation*}}
\newcommand{\eeql}{\end{equation*}}
\newcommand{\beqn}{\begin{eqnarray}}
\newcommand{\eeqn}{\end{eqnarray}}

\newcommand{\ua}{\uparrow}
\newcommand{\da}{\downarrow}

\begin{document}

\title{Lattice symmetry assisted second order topological superconductors and Majorana patterns}
\author{Xiao-Hong Pan}
\affiliation{School of Physics, Huazhong University of
Science and Technology, Wuhan, Hubei 430074, China}
\affiliation{Wuhan National High Magnetic Field Center, Huazhong University of
Science and Technology, Wuhan, Hubei 430074, China}
\author{Kai-Jie Yang}
\affiliation{School of Physics, Huazhong University of
Science and Technology, Wuhan, Hubei 430074, China}
\affiliation{Department of Physics, the Pennsylvania State University, University Park, PA, 16802, US}
\author{Li Chen}
\affiliation{School of Physics, Huazhong University of
Science and Technology, Wuhan, Hubei 430074, China}
\affiliation{Wuhan National High Magnetic Field Center, Huazhong University of
Science and Technology, Wuhan, Hubei 430074, China}
\author{Gang Xu}
\affiliation{Wuhan National High Magnetic Field Center, Huazhong University of
Science and Technology, Wuhan, Hubei 430074, China}
\affiliation{School of Physics, Huazhong University of
Science and Technology, Wuhan, Hubei 430074, China}
\author{Chao-Xing Liu}
\affiliation{Department of Physics, the Pennsylvania State University, University Park, PA, 16802, US}
\author{Xin Liu}
\affiliation{School of Physics, Huazhong University of
Science and Technology, Wuhan, Hubei 430074, China}
\affiliation{Wuhan National High Magnetic Field Center, Huazhong University of
Science and Technology, Wuhan, Hubei 430074, China}
\date{\today}

\begin{abstract}
We propose a realization of the lattice symmetry assisted second order topological superconductors with corner Majorana zero modes (MZM) based on two-dimensional topological insulators (2DTI). The lattice symmetry can naturally lead to the anisotropic coupling of edge states along different directions to the in-plane magnetic field and conventional s-wave pairings, thus leading to a single MZM locating at the corners for various lattice patterns. In particular, we focus on the 2DTI with D$_{3d}$ lattice symmetry and found different types of gap opening for the edge states along the armchair and zigzag edges in a broad range of parameters. As a consequence,  a single MZM exists at the corner between the zigzag and armchair edges, and is robust against weakly lattice symmetry broken. We propose to realize such corner MZMs in a variety of polygon patterns, such as triangles and quadrilaterals. We further show their potentials in building the Majorana network through constructing the Majorana Y-junction under an in-plane magnetic field. 
\end{abstract}

\maketitle

{\it Introduction -} Majorana zero modes (MZMs) in topological superconductors (TSCs), are extensively studied recently years \cite{Kitaev2001,Sau2010,Lutchyn2010,Oreg2010,Mourik2012,Deng2012,Rokhinson2012,Das2012,Wang2012,Churchill2013,Xu2014,Nadj-Perge2014,Chang2015,Sun2016,Albrecht2016,Wiedenmann2016,Bocquillon2016, Zhang2017,Zhang2018,Zhang2018b,Wang2018} because of their non-Abelian braiding statistics \cite{Kitaev2001,Ivanov2001,Nayak2008} and the potential application in topological quantum computation (TQC). Although the great experimental progress in several condensed matter platforms have led to the observation of zero bias conductance peak \cite{Mourik2012,Deng2012,Zhang2017,Zhang2018,Zhang2018b} and $4\pi$ Josephson effect \cite{Rokhinson2012,Wiedenmann2016,Bocquillon2016,Laroche2017,Deacon2017}, the deterministic evidence of the non-Abelian braiding statistics is still lacking for MZMs, which is essential for TQC. As the experimentally measurable braiding requires at least four MZMs, it is worth to search for new platforms that allow for the appearance of multiple MZMs. The recent studies of the second order topological states \cite{Zhang2013,Benalcazar2017,Fang2017,Peng2017,Benalcazar2017a,Langbehn2017,Song2017,Li2017,Ezawa2018c,Shapourian2018,Serra-Garcia2018,Peterson2018,Zhu2018,Ezawa2018b,Geier2018,Khalaf2018,Schindler2018,Ezawa2018a,
Liu2018,Wang2018c,Wieder2018,Schindler2018a,Ezawa2018d,Ezawa2018,Wang2018a,Wieder2018,Dwivedi2018,
Yan2018,Bultinck2018,Imhof2018,You2018,Wang2018b,Huang2018,
Hsu2018} have brought new insights of realizing MZMs. In contrast to conventional n-dimensional topological insulators (TIs), the second order TIs are characterized by topological protected gapless states in (n-2)-dimension. Particular for the two-dimensional second order TSCs \cite{Zhu2018,Liu2018,Wang2018a,Hsu2018,Dwivedi2018,Yan2018,Wang2018b}, the current proposals \cite{Yan2018,Wang2018a,Hsu2018} are based on s-, $s_{\pm}$- or $d$-wave superconductors and support at each corner a pair of MZMs protected by additional time-reversal or mirror symmetry. Since it is difficult to reveal non-Abelian statistics when MZMs are paired, it is thus more desirable to look for second order TSC with a single MZM at each corner. As there is no need of additional symmetry to
protect local single MZM, such system is also more robust against the environmental perturbation.

\begin{figure}[htbp]
  \centering
  \includegraphics[width=0.9\columnwidth]{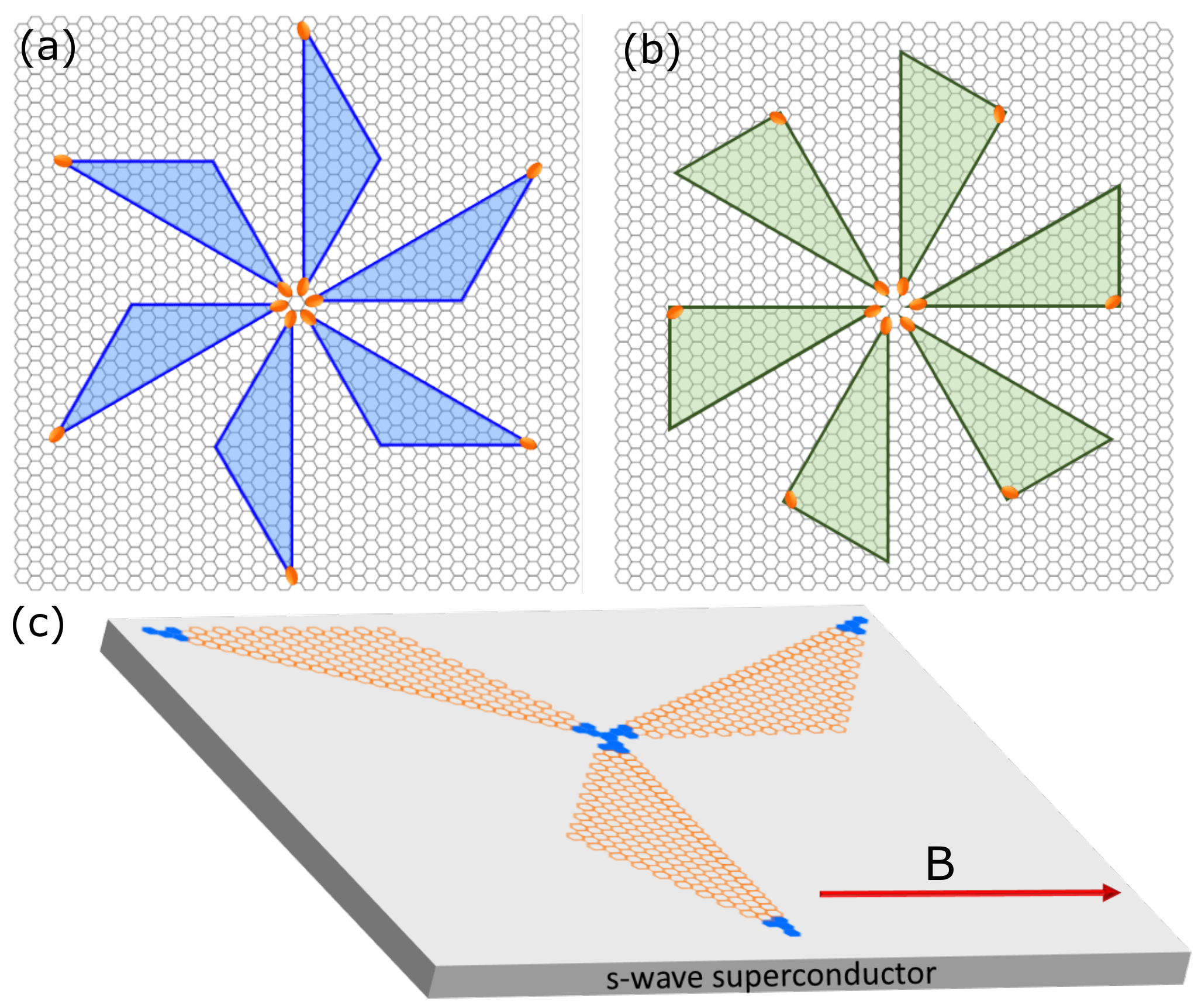}
  \caption{(a) The Majorana isosceles triangle in six orientations. (b) The Majorana right triangles in six orientations. Each Majorana triangle is equivalent with a Majorana nanowire along the six orientations. (c) Y-junction made from three Majorana isosceles triangles. The red arrow indicate the in-plane magnetic field direction.}
\label{pattern}
\end{figure}

In this work, we demonstrate the realization of a single MZM at certain corners of the 2DTIs with honeycomb lattice, in proximity to the conventional s-wave superconductor under an in-plane magnetic field. These 2DTIs includes the bismuthene and silicene, whose topological bands around Fermi surface are dominated by $\{p_x,p_y\}$ and $p_z$ orbitals respectively. Even in the quite general case for these different 2DTI models, we found that the helical edge states at the armchair edge still remain almost gapless while those at the zigzag edge are fully gapped. The anisotropic coupling of the edge states to the uniform in-plane magnetic field is due to the fact that the mirror symmetry at the edge states degenerate point is preserved along armchair edge but broken along the zigzag edge. This is in principle allowed because all of these models belong to $D_{3d}$ group, there are no symmetry operations which can transform the armchair edge to the zigzag edge. With further applying the uniform s-wave superconducting pairing term, a superconducting gap is open at the armchair edge while a magnetic gap occurs at the zigzag edge in a large parameter regime. As a result, we demonstrate a single MZM existing at the corner between the armchair and zigzag edges. We further identify two types of triangular patterns in the honeycomb lattices (Fig.~\ref{pattern}(a) and \ref{pattern}(b)), each of which supports two MZMs located separately at two corners of the triangle and is thus equivalent to a Majorana nanowire \cite{Kitaev2001,Sau2010,Lutchyn2010,Oreg2010}. We thus refer these patterns as Majorana triangles. More importantly, as the existence of MZMs in these patterns is independent of the in-plane magnetic field directions, we show that the Majorana triangles can have six orientations (Fig.~\ref{pattern}(a) and \ref{pattern}(b)) and thus can construct more complex Majorana network such as Y-junction under an uniform in-plane magnetic field (Fig.~\ref{pattern}(c)). Note that no additional symmetry other than the intrinsic particle-hole symmetry is required, and these MZMs are robust against the terms that weakly break the mirror symmetries, such as the Rashba SOC, weak disorders, and rotating the various Majorana patterns by a small angle. Finally we discuss the possible experimental realization.

\begin{figure}[htbp]
  \centering
  \includegraphics[width=1.0\columnwidth]{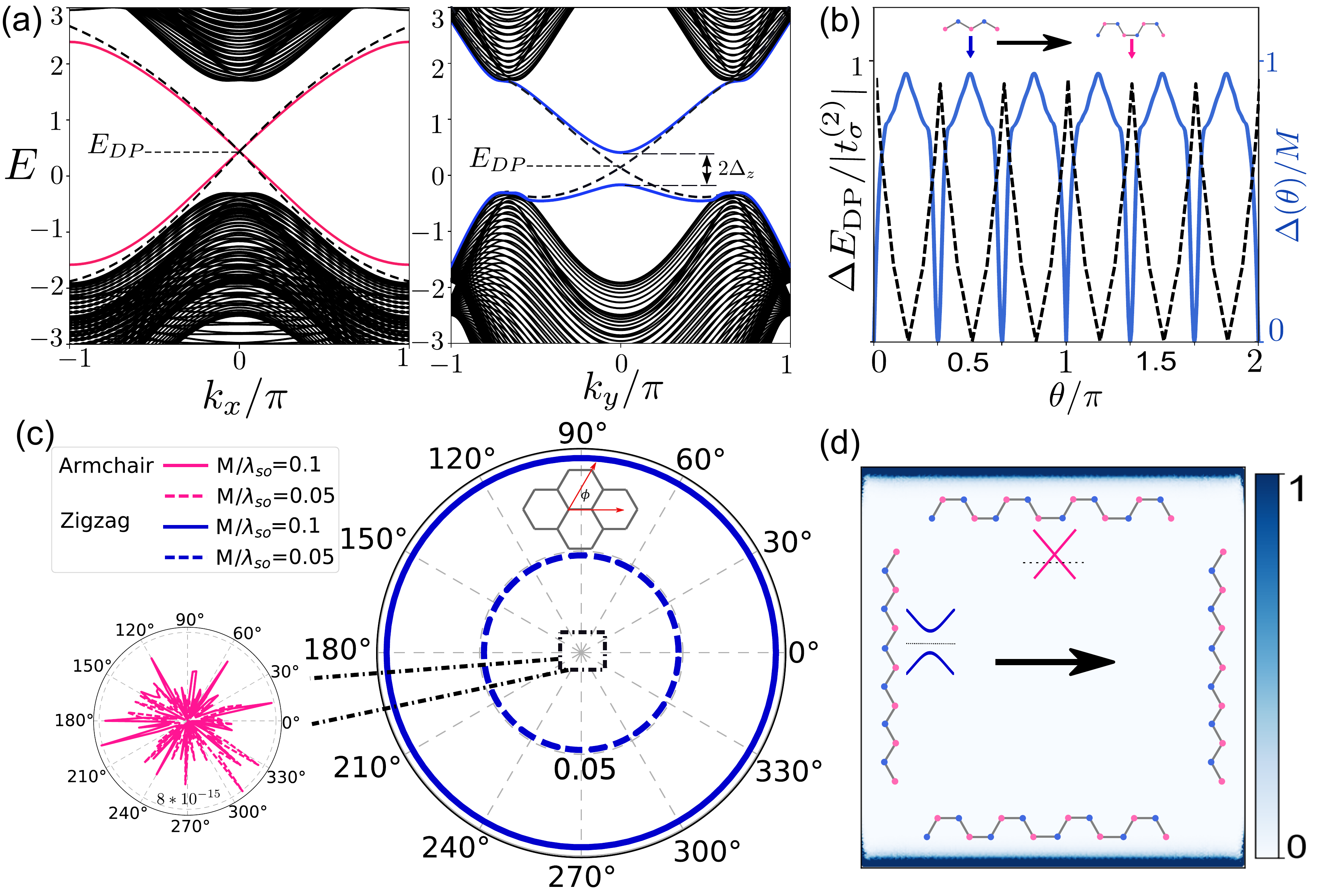}
  \caption{(a) The electronic band structures along armchair (left panel) and zigzag (right panel) edges. The solid and dashed curves represent the edge state dispersions with and without in-plane magnetic field. (b) The relative energies of Dirac points $\Delta E_{DP}=E_{\rm DP}(\theta)-E_{\rm DP}(\pi/2)$(the black dashed curve) and the edge state gap (the blue curve) as a function of the edge cut directions with the the magnetic field along $x$ direction (black arrow). (c) The gap at the zigzag (blue curves) and armchair edges (red curves) as a function of in-plane magnetic field directions with different amplitudes. (d) The wave function plot with its eigenvalue closest to zero. The black arrow indicates the magnetic field direction.}
\label{FM}
\end{figure}

{\it Anisotropic edge states gap -} To demonstrate the experimental accessibility of our proposal, we start from the Zhang-Li-Wu (ZLW) model \cite{Zhang2014} for single layer bismuth grown on SiC substrate, which has been experimentally reported to have large quantum spin Hall gap around 0.435eV \cite{Reis2017} and one-dimensional edge states \cite{Stuehler2019}. However our results are also applied to other 2DTI such as Kane-Mele model for graphene \cite{Kane2005,Kane2005a} and Liu-Jiang-Yao model \cite{Liu2011a} for silicene with $p_{\rm z}$. The ZLW model with $\{p_x, p_y\}$ orbitals forming the 2DTI bands takes the form \cite{Zhang2014,Li2018}
\beqn\label{Ham-1}
H &=& \sum_{\langle ij \rangle}t_{\sigma}^{(1)}p_{i,\bm{a}_{ij}}^{\dag}p_{j,\bm{a}_{ij}}+t_{\pi}^{(1)}{p'}_{i,\bm{a}_{ij}}^{\dag}p'_{j,\bm{a}_{ij}} \nonumber \\
&+&  \sum_{\langle\langle ij \rangle\rangle} t_{\sigma}^{(2)} p_{i,\bm{b}_{ij}}^{\dag}p_{j,\bm{b}_{ij}}+t_{\pi}^{(2)}{p'}_{i,\bm{b}_{ij}}^{\dag}{p'}_{j,\bm{b}_{ij}}\nonumber \\
  &-&\sum_{i,s}i \lambda_{\rm so} p_{i,x}^{\dag}s_z p_{i,y}+H.c. 
\eeqn
where $t_{\sigma(\pi)}^1,t_{\sigma(\pi)}^{2},\lambda_{\rm so}$ are the usual $\sigma(\pi)$ bond strengths between nearest neighbor sites, next-nearest neighbor sites, the intrinsic  SOC strengths, respectively, $p_{i,\bm{a}_{ij}}$ ($p_{i,\bm{b}_{ij}}$) and $p'_{i,\bm{a}_{ij}}$ ($p'_{i,\bm{b}_{ij}}$) are the projections of $\{p_{x},p_{y}\}$ orbitals parallel and perpendicular to the bond direction $\bm{a}_{ij}$ ($\bm{b}_{ij}$) for the first (second) nearest hopping respectively \cite{supp}, $s_z=\pm$ refers to spin-up and spin-down, $\langle ... \rangle$ and $\langle\langle...\rangle\rangle$ are the summations for the nearest and next-nearest neighbors respectively. In the rest of this work, we take $t_{\sigma}^{(1)}=2$eV, $t_{\pi}^{(1)}=-0.21$eV, $t_{\sigma}^{(2)}=-0.15$eV, $t_{\pi}^{(2)}=-0.05$eV, $\lambda_{\rm so}=0.435$eV according to the Ref.\onlinecite{Reis2017} 
The system has time-reversal symmetry protected gapless helical edge states (black dashed curves in Fig.~\ref{FM}(a)). As the second nearest neighbor hopping breaks the \lq\lq{}particle-hole\rq\rq{} symmetry (the symmetric band dispersion between conduction and valence bands), the energies of Dirac points, $E_{\rm DP}$, at armchair and zigzag edges are different which is normally the case for the real situation \cite{Reis2017}. We employ the python package Kwant \cite{Groth2014} to perform the numerical calculations in this work. The $E_{\rm DP}$ shows a C$_3$ rotational symmetry (black dashed curves in Fig.~\ref{FM}(b)).

We now consider applying a uniform in-plane magnetic field into the system. The Zeeman splitting under an in-plane magnetic field in both bismuthene and silicene systems takes the form  $M s_{\parallel}$ \cite{supp}, which in general is expected to open a gap, referred to as Zeeman gap, for the helical edge states along any edge directions because it breaks time-reversal symmetry and couples the states with opposite spin along the $z$ direction. Here $s_{\parallel}$ implies the spin along the in-plane magnetic field direction. We first focus on the armchair and zigzag edges and fix the magnetic field direction along the $x$ direction. We find that the helical edge states at zigzag edge acquires a finite gap $\Delta_{z}$, with the amplitude approximately equal to the Zeeman splitting energy. On the other hand, the Zeeman gap at armchair edge, $\Delta_{\rm a}$, is very small, the ratio $\Delta_{a}/M < 10^{-2}$ which are similar with the previous study in either silicene or bismuthene without breaking \lq\lq{}particle-hole\rq\rq{} symmetry \cite{Kuzmanovski2016,Dominguez2018}. Remarkably, we find the quasi-metallic state at armchair edge is also robust against breaking the \lq\lq{}particle-hole\rq\rq{} symmetry. The Zeeman gap and its gap center are plot in terms of the edge direction given the magnetic field along $x$ direction and shows highly anisotropy with a periodicity of $\pi/3$ (Fig.~\ref{FM}(b)) which reflects the $C_3$ symmetry of the Hamiltonian (Eq.~\eqref{Ham-1}). We further explore whether the magnetic field direction affects our results.  In Fig.~\ref{FM}(c), we focus on the zigzag (armchair) edge along the $y$ ($x$) direction and plot the Zeeman gaps in the blue (red) curves as a function of the magnetic field direction. For all directions, the Zeeman gap at zigzag edge takes the value around the Zeeman splitting energy while the armchair edge remains almost gapless. Meanwhile, the energies of the Dirac point along the two edges are also not changed when rotating magnetic field. Thus when the zigzag edges are insulating, the armchair edges always behave like a one-dimensional (1D) single channel metallic wire (Fig.~\ref{FM}(d)) regardless of the in-plane magnetic field direction. 

\begin{figure}[htbp]
  \small
  \centering
  \includegraphics[width=1.0\columnwidth]{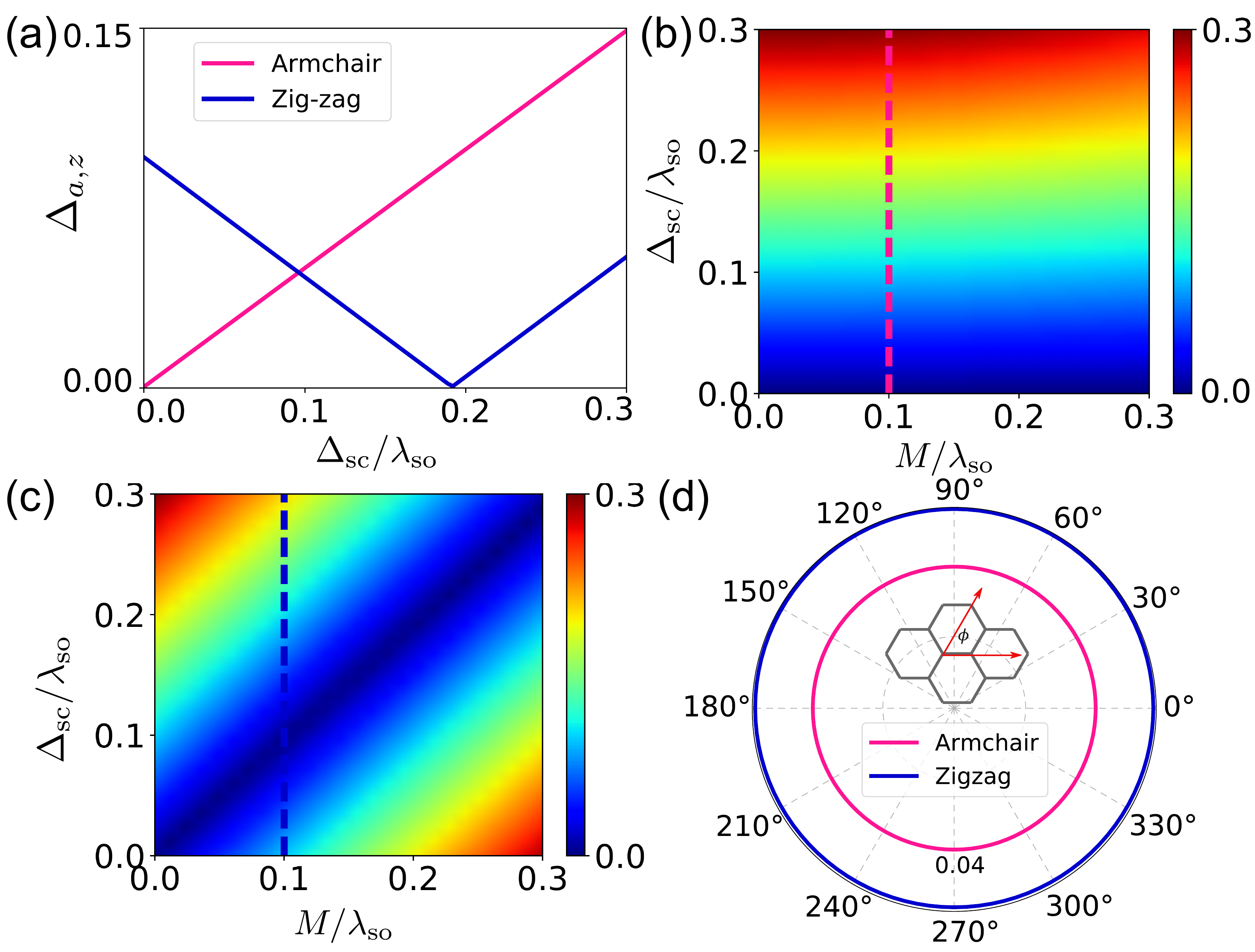}\\
  \caption{(a) The gaps for the zigzag and armchair edges as a function of the superconducting gap $\Delta_{\rm sc}$ with the in-plane magnetic field, $M/\lambda_{\rm so} = 0.1$, along $x$ direction. (b) and (c) The color plot of the gap at the armchair and zigzag edges respectively under the in-plane magnetic field and superconducting amplitudes.  (d) The edge state gap as a function of the in-plane magnetic field direction with $M/\lambda_{\rm so} = 0.1$ and $\Delta_{\rm sc}/\lambda_{\rm so} = 0.04$. For this plot, we choose the armchair and zigzag edges along $x$ and $y$ directions respectively.}
\label{SC}
\end{figure}

We further apply an uniform conventional s-wave and spin-singlet superconducting gap function with gap amplitude $\Delta_{\rm sc}$ into system while keeping the in-plane magnetic field. For simplicity, we take $\mu$ to be the Zeeman gap center at zigzag edges but our results generally remain valid for varying $\mu$ \cite{supp}. In Fig.~\ref{SC}(a), we plot the edge state gaps, $\Delta_{a}$ and $\Delta_{z}$, as a function of the superconducting gap $\Delta_{\rm sc}$ with fixed $M$ for the armchair (red curves) and zigzag (blue) edges, respectively. For the armchair edge, the previous metallic edge states are gapped with $\Delta_{\rm a} \approx \Delta_{\rm sc}$. For the zigzag edge, the gap first decreases to zero at $M \approx \Delta_{\rm sc}$ and then reopens with further increasing $\Delta_{\rm sc}$ (Fig.~\ref{SC}(a)). This implies that the edge states at zigzag edge undergoes a phase transition, from Zeeman dominated gap states to superconductivity dominated gap states. In Fig.~\ref{SC}(b) and \ref{SC}(c), we plot the edge states gap at armchair edge and zigzag edge respectively as a function of superconducting gap amplitude $\Delta_{\rm sc}$ and the Zeeman splitting energy $M$. For armchair edge, the gap remains finite as long as $\Delta_{\rm sc}$ is finite and thus the gap $\Delta_{\rm a}$ is dominated by superconductivity regardless of the strength of magnetic field. For zigzag edge, the gap is closed approximately when $\Delta_{\rm sc}$ and $M$ are equal. We further plot the gaps at  two edges as a function of the magnetic field direction and found that the gap amplitudes on two edges are independent of the magnetic field directions (Fig.~\ref{SC}(d)). These results hold for the boundary along all of the armchair and zigzag directions which exhibit similar gap amplitudes for the same type of edges. Thus, roughly in the range $0<\Delta_{\rm sc} < M$, the armchair and zigzag edges are in two topologically different phases. This implies that a single MZM exists at the corner between these two types of edges.

\begin{figure}[htbp]
  \centering
  \includegraphics[width=1.0\columnwidth]{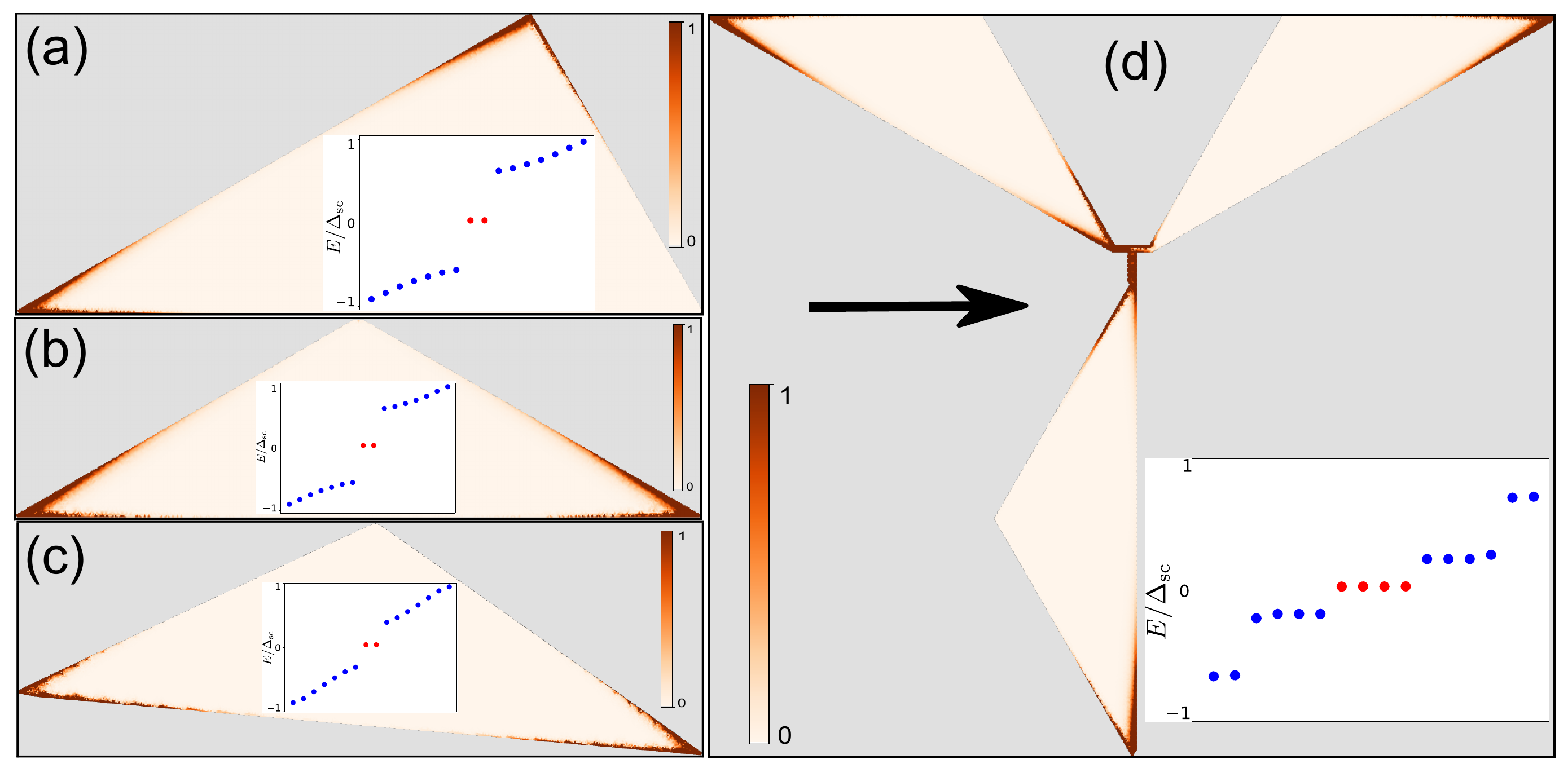}
  \caption{(a) and (b) The density plot of MZMs in Majorana isosceles and right triangle. (c) Majorana triangle with the edges have $5^{\circ}$ deviation from the armchair and zigzag edges respectively. (d) Y-junction made from three Majorana isosceles triangles. The black arrow indicate the in-plane magnetic field direction. The insets plot the energies of the eigenstates.  }
    \label{density}
\end{figure}

{\it Majorana patterns-} As discussed above, our theory has indicated the existence of MZMs at the corner between the armchair and zigzag edges. Below, we will consider the realization of these corners in the sample patterns of polygons, particularly triangle patterns. As the whole system only allows for even number of MZMs, the triangle can only support up to two MZMs at the two of its three corners. For the honeycomb lattice, the angles between the armchair and zigzag edges can take the values $\pi/6$, $\pi/2$ and $5\pi/6$ and while those between the same type of edge are $\pi/3$ and $2\pi/3$. We hope that three edges of the triangle are either armchair or zigzag boundary and identify two triangle configurations for this condition. One is an obtuse isosceles triangle with the obtuse interior angle of $2\pi/3$ (Fig.~\ref{pattern}(a))  and the other is a right triangle with one acute interior angle of $\pi/6$ (Fig.~\ref{pattern}(a)). Due to the $C_{3}$ rotational symmetry of the honeycomb lattice, there are six orientations for each triangle configuration as shown in Fig.~\ref{pattern}(a) and \ref{pattern}(b). We then calculate the eigenvalues of the system with these two triangle configurations and found that in the range $0<\Delta_{\rm sc}<M$, each of them support two zero modes, shown in the insets of the Fig.~\ref{density}(a) and \ref{density}(b). The density plots of these two zero modes in the two triangular patterns (Fig.~\ref{density}(a) and (b)) show that  they separately locate at the two of three corners with the interior angles of $\pi/6$ for the isosceles triangle and with the interior angles of $\pi/6$ and $\pi/2$ for the right triangle. We thus dub these two triangle configurations as Majorana triangles.  As these Majorana triangles host only two MZMs, they are topologically equivalent to the Majorana nonawire \cite{Kitaev2001,Sau2010,Lutchyn2010,Oreg2010}. Importantly, because the applied in-plane magnetic field direction will not affect the gaps at armchair and zigzag edges, all the Majorana triangles, on these six orientations can have MZMs under the same in-plane magnetic field. Note that the local single MZM is topologically stable and does not require additional symmetry, we rotate the Majorana triangle by $5^{\circ}$ and the spacially separated MZMs are still stable even there is no symmetries at each edge. All these features lead the advantage to realize more complex Majorana structures. In Fig.~\ref{density}(d), we construct the Y-junction  which is proposed to realize Majorana braiding \cite{Clarke2011,Sau2011,Heck2012,Hyart2013,Wu2014}. The density plot and the eigenvalues calculations show that there are four MZMs in these constructions are well separated from each other. The additional two modes with the eigenenergies closer to zero than other excited states are from the coupled Majorana bound states at the center of the Y-junction. The color bar with logarithmic scale shows they are well separated from the four MZMs in the energy space. Importantly, it should be noted that this Y-junction is realized under the uniform in-plane magnetic field. Note that the MZM at each corner is protected by topology, the slightly symmetry broken by Rashba SOC does not affect the robustness of the MZM at all. 

{\it Magnetic gap at the armchair edge -} The anisotropic gap at the armchair and zigzag edges plays the essential role for obtaining Majorana corner states and we will here analyze the gap anisotropy based on the edge theory. We found although bismuthene and silicene have different Hamiltonian, the two branches of the helical edge states cross at $k_x=0$ along armchair edge (Fig.~\ref{FM}(a)) \cite{supp}. Meanwhile the next nearest neighbor hopping term become a constant after the projection. Thus taking $k_x = 0$, the 1D Hamiltonian along $y$ direction at basis $(p_{-,\ua}^{A},p_{+,\ua}^{B},p_{-,\da}^{A},p_{+,\ua}^{B})^{T}$ takes the form \cite{supp}
\beqn\label{Ham-2}
H_0 &= &\frac{1}{2}t_{m}^{(1)}\left[1-\sqrt{3}\sin(i\frac{1}{2}\partial_y)-\cos(i\frac{1}{2}\partial_y)\right]s_{0}\sigma_{x}   \nonumber\\
&+&\lambda_{\rm so}s_{z}\sigma_{z} , \nonumber\\
H\rq{}&=& \frac{\sqrt{3}}{4} t_{m}^{(1)}k_{x}s_{0}\sigma_{y} -\frac{3}{2}t_{p}^{(2)}s_{0}\sigma_{0}+ Ms_{\parallel}\sigma_{0}.
\eeqn
with $p_{\pm}=p_{x}\pm i p_{y}$, $t_{m}^{(1)}=(t_{\sigma}^{(1)}-t_{\pi}^{(1)}), t_{p}^{(2)}=(t_{\sigma}^{(2)}+t_{\pi}^{(2)})$, $\sigma$ the pauli matrix acting in sublattice space. Here the next-nearest hopping and in-plane Zeeman terms are treated as the perturbation Hamiltonian. As the zero order Hamiltonian (Eq.~\eqref{Ham-2}) in the sublattice space only contains $\sigma_x$ and $\sigma_z$,  the edge states for the semi-infinite system with $y \in (-\infty,0)$ generally take the form
\beqn\label{wf-1}
&&\Psi_{n=1,2}(y) = N \sin(\alpha y) e^{\beta y} \chi_{n=1,2}, \nonumber \\
&&\chi_{1} = \ket{\ua} \otimes \ket{\sigma_{y}=1}, \ \ \chi_{2} = \ket{\da} \otimes \ket{\sigma_{y}=-1},
\eeqn
with $\alpha$ and $\beta$ the wave vector and the decay rate of the edge states. Here $\chi_1$ and $\chi_2$ are time-reversal partners. Noted that this form of the helical edge state is not by accident, but enforced by the mirror symmetry. This is because the nearest neighbor hopping term conserves spin and thus can be only proportional to $\sigma_x s_0$ and $\sigma_y s_0$. Meanwhile, along the armchair edge, it also respects the mirror-x symmetry ($\hat{M}_x=i s_x\sigma_x$), so that the first term on the right-hand side of Eq.~\eqref{Ham-2} can be only proportional to $\sigma_x s_0$ \cite{supp}. In addiction, the SOC term is always proportional to $\sigma_z s_z$. Thus the helical edge states at $k_x=0$ must take the form of Eq.~\eqref{wf-1} \cite{supp}. We then project the 1D Hamiltonian along $y$ direction with finite $k_x$ \cite{supp} into the two dimensional basis $\Psi_{n}e^{i k_x x}$ and get the effective edge state Hamiltonian \cite{supp}
\beqn\label{edge-Ham}
H_{\rm edge}= -\frac{\sqrt{3}}{4}t_{m}^{(1)} k_x\tilde{s}_z -\frac{3}{2}t_{p}^{(2)},
\eeqn
with $\tilde{s}$ acting on the $(\Psi_1,\Psi_2)^{\rm T}$ basis. Note that $\bra{\Psi_i} M\sigma_0 s_{\parallel} \ket{\Psi_j} = 0$ for all $i=1,2$ and $j=1,2$, the in-plane magnetic Zeeman term completely vanishes in the effective edge Hamiltonian. We found that for the edge states in Eq.~\eqref{wf-1}, only the anti-ferromagnetic term such as $\sigma_z s_{\parallel}$ can directly open a gap while the ferromagnetic term $\sigma_0 s_{\parallel}$ can not. The in-plane magnetic field lead to ferromagnetic-like Zeeman splitting \cite{supp} and thus cannot open a gap at the armchair edge. On the other hand, the anti-ferromagnetic term can come from the magnetic field fluctuation. Note that this analysis holds for all the armchair edges with the in-plane ferromagnetic or anti-ferromagnetic terms in all directions. Although anti-ferromagnetic term can open gap in both armchair and zigzag edges, the Majorana corner states still remains robust because when the chemical potential inside the zigzag edge state gap, it is away from the $E_{\rm DP}$ at armchair edge so that the armchair edge are always superconducting. Thus, the Majorana corner states remain even in the presence of the in-plane anti-ferromagnetic term \cite{supp} and thus robust against the inhomogeneity of the magnetic field. We also find that the weak spin-independent disorder will not affect the wave function forms of $\chi_1$ and $\chi_2$. Thus our results remain valid against weak spin-independent disorder \cite{supp}.

{\it Conclusion and discussion-} In conclusion, we propose to realize the second order TSC in the D class and the corner MZMs based on 2DTI under a uniform in-plane magnetic field and in proximity to s-wave superconductors. Our scheme is shown with the realistic bismuthene model but also valid for other 2DTI model such as silicene, germanene and stanene, and may have the advantage in constructing Majorana networks. For the mono-layer NbSe$_2$ superconductor, the superconducting gap is about 0.5meV and the in-plane critical field can be as large as 27T \cite{Xi2015}. For the in-plane magnetic field of 10T with the g-factor $g=2$, the Zeeman splitting energy is about 1.2meV. As the single MZM is topologically robust against the local perturbation, we show that our results hold even in the presence of various perturbations. So our proposal maybe realized under the reasonable material parameters. 
 
\section*{Acknowledgement}

We would like to thank Yugui Yao, Cheng-Cheng liu, Jinhua Gao, Aiyun Luo and Xiong-Jun Liu for fruitful discussions. This work is supported by National Key R\&D Program of China (Grant No. 2016YFA0401003) and NSFC (Grant No.11674114).


%

\end{document}